\documentclass[aps,11pt,showpacs,superscriptaddress,amsfonts,amssymb,amsmath]{revtex4}

\usepackage{graphicx}

\begin{document}

\title{Velocity requirements for causality violation}

\medskip

\author{G.\ Modanese \footnote{Email address: giovanni.modanese@unibz.it}}

\affiliation{Free University of Bolzano, Faculty of Science and Technology, Bolzano, Italy \medskip}

\date{June 16, 2014}

\linespread{0.9}

\begin{abstract}

\medskip

We re-examine the ``Regge-Tolman paradox'' with reference to some recent experimental results. It is straightforward to find a formula for the velocity $v$ of the moving system required to produce causality violation. This formula typically yields a velocity very close to the speed of light (for instance, $v/c > 0.97$ for X-shaped microwaves), which raises some doubts about the real physical observability of the violations. We then compute the velocity requirement introducing a delay between the reception of the primary signal and the emission of the secondary. It turns out that in principle for any delay it is possible to find moving observers able to produce active causal violation. This is mathematically due to the singularity of the Lorentz transformations for $\beta \to 1^-$. For a realistic delay due to the propagation of a luminal precursor, we find that causality violations in the reported experiments are still more unlikely ($v/c > 0.989$), and even in the hypothesis that the superluminal propagation velocity goes to infinity, the velocity requirement is bounded by $v/c > 0.62$. We also prove that if two macroscopic bodies exchange energy and momentum through superluminal signals, then the swap of signal source and target is incompatible with the Lorentz transformations; therefore it is not possible to distinguish between source and target, even with reference to a definite reference frame.

\medskip


\end{abstract}


 \maketitle

\section{Introduction}
\label{intro}

Phenomena of electromagnetic wave propagation with superluminal group velocity have been observed in several laboratories in the last years and can be collected in two categories: (a) evanescent waves, and (b) Bessel beams of so-called ``X-shaped waves''.

Concerning the first category, superluminal effects for evanescent waves  have been demonstrated in tunnelling experiments in both the optical domain and microwaves range \cite{eva1,eva2,eva3,eva4}; these effects can be revealed, however, only over short distances, typically a few centimetres for microwaves (the most favourable case).

Concerning the second category, Mugnai et al.\  have demonstrated  the superluminal propagation of localized microwaves over a distance of 1 m or more \cite{Mugnai}. The field of the beam can be considered as formed by the superposition of pairs of X-shaped plane waves. These move with velocity approximately up to 25\% in excess of light speed. A similar experiment was performed in the optical range \cite{Opt1,Opt2}, but a clear observation of superluminal propagation was impossible in that case. More recently, Missevitch et al.\ demonstrated anomalously small retardation of bound UHF electromagnetic fields within about the half of the near zone size \cite{miss}.

Several papers discuss the issue of signal transmission in experiments of this kind (see \cite{break} for a recent review and references). The question is, if superluminal propagation effects can be used to convey information at superluminal speed, and the answer is generally that they can't, though it also depends on what is exactly meant by a signal. A typical argument is that waves with superluminal group velocity are always accompanied by a ``precursor wave'' propagating at light speed. Some authors speculated, however, that in certain cases the superluminal wave could overtake the precursor. Other authors argued that the violation of causality by true superluminal signals is only apparent and could be avoided through the so-called Feynman-Stuckelberg tachyon reinterpretation principle.

In this work we reconsider the relation between superluminal propagation phenomena and the violation of causality. We distinguish between ``passive'' causality violation (the possibility to {\em observe} a cause-effect inversion in a suitable moving system) and ``active'' causality violation (the possibility that an effect triggers the disabling of its own cause). For both cases it is straightforward to compute the velocity that the moving system must have in order to make the causality violation possible. For the case of active violation, however, we introduce the possibility that there is a dead time in the re-emission of the superluminal signal. We include in the calculation also extreme cases, with large superluminality and/or large delays. Finally, we analyse the exchange of tachyons between two macroscopic sources from the point of view of a moving system and prove a property of generalized Lorentz invariance which  prevents any distinction between source and target. 

It is known that quantum field theories with tachyons are plagued by instabilities; note that throughout this work we consider tachyons not as fundamental particles, but only as a possible formal representation of superluminal signals.

\section{Passive and active causality violation}
\label{sec:3}

By ``passive'' causality violation we simply mean the fact that in suitable moving systems the processes of generation and detection of a superluminal signal with $V>c$ are inverted in time. A direct application of Lorentz transformations shows that in order to see this causality inversion the moving observer must have a velocity $v > c^2/V$.

\begin{figure}
  \includegraphics[width=9cm,height=6cm]{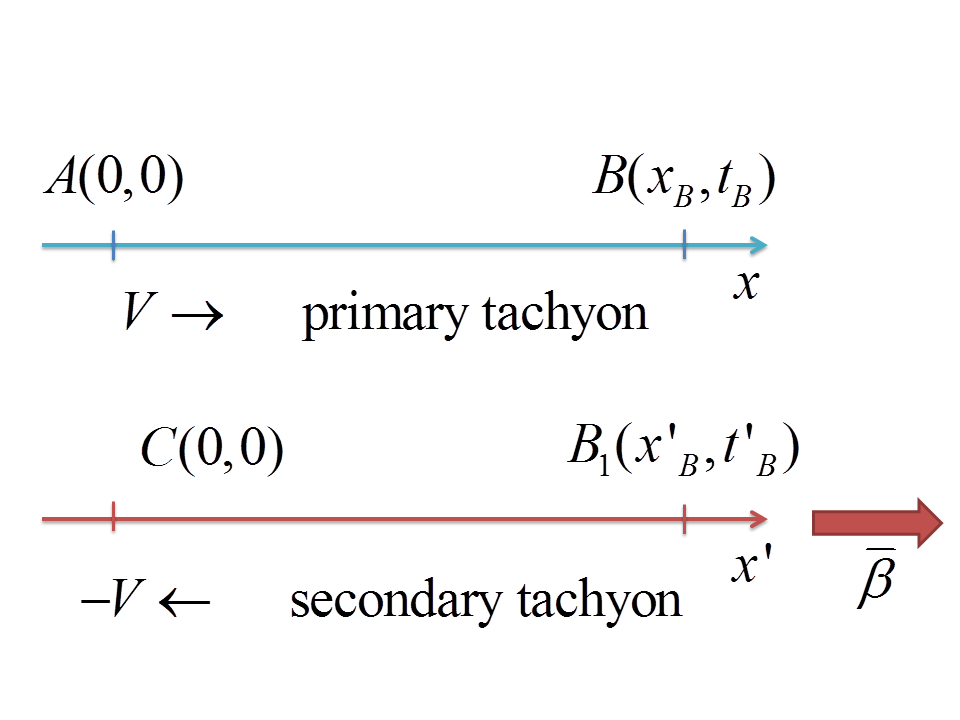}
\caption{Regge-Tolman paradox: in the laboratory system a primary tachyon is emitted in $A$ and detected in $B$. In a system moving with velocity $\bar{\beta}c$ with respect to the laboratory it is possible to emit a secondary tachyon (event $B_1$) which reaches the origin of the laboratory system (event $C$) at the same time of the primary emission. We suppose initially that $B$ and $B_1$ coincide (no delay between the reception of the primary tachyon in $B$ and the emission of the secondary tachyon in $B_1$). The velocity requirement for $\bar{\beta}$ is found to be given by eq. (\ref{eq9}), namely $\bar \beta  = 2Vc/({c^2} + {V^2})$ .}
\label{Tolman}       
\end{figure}

Let us now look at the velocity requirements for ``active'' causality violation, in the form of the so-called ``tachyon anti-telephone" first discussed by Tolman (\cite{Benford} and ref.s). The argument is written down in detail, in order to generalize it later to the case of re-emission with delay. We admit the existence of some device which emits particles with propagation velocity $V>c$ (``tachyons"), and also the existence of efficient detectors for such particles. We want to use these particles for communication in spacetime. Let us also suppose, for a start, that the emission and detection lags can be disregarded, so that the communication timing is determined only by the propagation intervals. We ask if it is possible not only to observe a causality violation in some reference system, but to actively ``send secondary superluminal signals back in time" and switch off the source of the primary signals even before their emission (Fig.\ \ref{Tolman}). This is an impressive logical evidence of causality violation. What are the velocity requirements?

Consider two reference systems: the first one is the ``laboratory system", with coordinates $(x,t)$, the second one the ``moving system", with coordinates $(x',t')$, which is moving with respect to the laboratory system in the positive $x$ direction, with velocity $v=\beta c$. Suppose that the origins of the two systems coincide at the initial time, that is, the origin $(0,0)$ denotes the same event in the two reference systems. In the laboratory system there is a tachyon emitter at the origin $x=0$; at the time $t=0$ this emits a primary tachyon which travels in the positive $x$ direction and is detected at time $t_B$ by a receiver placed at $x=x_B$. Therefore the event $A(0,0)=A'(0,0)$ is the emission of the particle, and $B(x_B,t_B)$ its reception. We have $t_B=x_B/V$, with $V>c$. In the moving system the coordinates of the reception are $B'(x'_B,t'_B)$ and are related to the coordinates $x_B$, $t_B$ by the Lorentz transformation
\begin{equation}
\begin{array}{l}
x{'_B} = \gamma {x_B} - \beta \gamma c{t_B}\\
t{'_B} = \gamma {t_B} - \beta \gamma \frac{{{x_B}}}{c}
\end{array}
\label{eq1}
\end{equation}
Now suppose that just when the laboratory detector is hit by the ``primary" tachyon, a moving emitter (which is at rest in the moving system) is near $B$ and sees the detection with negligible delay. The moving emitter then emits a ``secondary" tachyon, in the negative $x'$ direction. We call the secondary emission Event $B_1$; it coincides with $B$ in the absence of any emission delay, and for now let us suppose that this is the case.

The event $C$ is the transit of the secondary tachyon near the origin of the laboratory system, where the primary emitter is placed. This event has coordinates $(0,t_C)$ in the laboratory system and $(x_C',t_C')$ in the moving system. Finally suppose that in the moving system several detectors are placed along the path of the secondary tachyon; if one of these detectors is near the primary emitter and receives the secondary tachyon, then it is programmed to disable the primary emitter.

Our task is to check under which conditions the time of the event $C$, in the laboratory system, is positive, zero or negative. If $t_C>0$, then there is no causality violation, because the switching-off of the primary emitter, ultimately caused by the primary emitter itself, occurs after the primary emission. On the contrary, if $t_C\leq 0$, then we are confronted with active causality violation. It is straightforward to prove, using diagrams representing the spacetime trajectories of the primary and secondary tachyons, that $t_C$ can become negative if the velocity of the moving system is sufficiently close to $c$. Here we want to compute the exact ``critical" value $\bar{\beta}$  of the $\beta$ parameter for which $t_C=0$. For $\beta  > \bar \beta $ active causality violation occurs.

\begin{figure}
  \includegraphics[width=9cm,height=6cm]{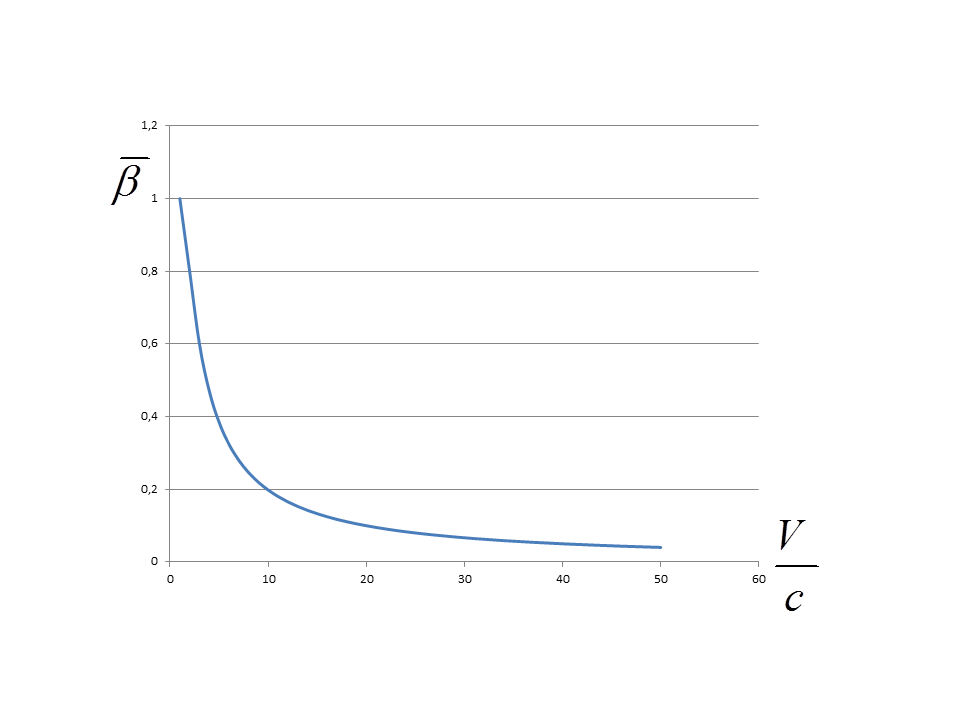}
\caption{Velocity requirement for active causality violation (eq.\ (\ref{eq9})). }
\label{beta}       
\end{figure}

We denote with $\tau$ the time elapsed, in the moving system, since the event $B_1$. At the time $(t'_B+\tau)$, in the moving system, the secondary tachyon emitted in $B_1$ is at the position
\begin{equation}
x' = x{'_B} - V\tau
\label{eq2} 
\end{equation}
In the laboratory system this position is transformed to
\begin{equation}
x = \gamma x' + \beta \gamma c(t{'_B} + \tau )
\label{eq3} 
\end{equation}
Setting $x=0$ and solving simultaneously (\ref{eq2}) and (\ref{eq3}), we find the time $\tau_C$ that the secondary tachyon takes (for a fixed $\beta$), to reach the point of the primary emission. The result is
\begin{equation}
{\tau _C} = \frac{{x{'_B} + \beta ct{'_B}}}{{V - \beta c}}
\label{eq4} 
\end{equation}
and the instant when this happens is
\begin{equation}
t{'_C} = t{'_B} + {\tau _C}
\label{eq5} 
\end{equation}
Now, by imposing that $t'_C=0$, we find the critical value of $\beta$ such that the event $C$ (secondary tachyon arrives at the emission location of the primary tachyon) occurs at the same time as the primary emission, i.e. $(x'_C, t'_C)=(0,0)$. (Note that we have set $x'_C=0$ already, after eq.\ (\ref{eq3})). From (\ref{eq4}), (\ref{eq5}) we obtain
\begin{equation}
\frac{{x{'_B} + \bar \beta ct{'_B}}}{{V - \bar \beta c}} + t{'_B} = 0
\label{eq6} 
\end{equation}
and hence
\begin{equation}
x{'_B} + Vt{'_B} = 0
\label{eq7} 
\end{equation}
Transforming into the laboratory system:
\begin{equation}
\bar \gamma {x_B} - \bar \beta \bar \gamma c{t_B} + V\left( {\bar \gamma {t_B} - \bar \beta \bar \gamma \frac{{{x_B}}}{c}} \right) = 0
\label{eq8} 
\end{equation}
and finally, recalling that $x_B=Vt_B$, we obtain the desidered velocity requirement (graph in Fig.\ \ref{beta}):
\begin{equation}
\bar \beta  = \frac{{2V}}{{c + \frac{{{V^2}}}{c}}}
\label{eq9} 
\end{equation}
For slightly superluminal signals $(V=c(1+\varepsilon))$ one finds $\bar \beta \simeq 1-\varepsilon^2/2$. For instance, for the X-shaped waves of Mugnai et al.\ \cite{Mugnai}, superluminality of 7 \% leads to $\bar \beta \simeq 99.8$ \%, while superluminality of 25 \% leads to $\bar \beta \simeq 97$ \%.

\section{Modification of the velocity requirements in the presence of a dead time}
\label{modif}

Let us now suppose that the emission of the second tachyon is delayed with respect to the detection of the primary tachyon. Between the events $B$ and $B_1$ there will be a certain dead time $\Delta t$ (measured in the moving system), and eq.\ (\ref{eq2}) must be modified as follows: when $\tau<\Delta t$ the second tachyon is not present, while when $\tau \geq \Delta t$ the position of the second tachyon is
\begin{equation}
x' = x{'_B} - V\left( {\tau - \Delta t} \right)
\label{eqR1} 
\end{equation}
In the laboratory system this transforms to
\begin{equation}
x = \gamma x' + \beta \gamma c\left( {t{'_B} + \tau } \right)
\label{eqR2} 
\end{equation}
Setting as before $x=0$ and solving the system between (\ref{eqR1}) and (\ref{eqR2}) one obtains the time $\tau_C$ when the second tachyon arrives in $C$:
\begin{equation}
{\tau _C} = \frac{{x{'_B} + \beta ct{'_B} + V\Delta t}}{{V - \beta c}}
\label{eqR3} 
\end{equation}
(This obviously reduces to (\ref{eq4}) when $\Delta t=0$.) Now we set $t{'_C} = t{'_B} + {\tau _C} = 0$  and find an equation for the critical value $\bar \beta$  necessary for active causality violation with delay. After some algebra we find the irrational equation
\begin{equation}
2 - \bar \beta \left( {\frac{c}{V} + \frac{V}{c}} \right) + \frac{{\Delta t}}{{{t_B}}}\sqrt {1 - {{\bar \beta }^2}}  = 0
\label{eqR4} 
\end{equation}
(Again, note that this reduces to (\ref{eq9}) for $\Delta t=0$.) Define the following parameters, supposed to be known:
\begin{equation}
\begin{array}{l}
r = \frac{c}{V} + \frac{V}{c}\\
s = \frac{{\Delta t}}{{{t_B}}}
\end{array}
\label{eqR5} 
\end{equation}
The parameter $r$ depends on the propagation velocity, while $s$ depends on the ratio between the delay in the secondary emission and the flight time of the primary tachyon. (Note that this flight time and the distance from the emitter have no influence on the requirement for causality violation, when there is no delay.) In a typical situation where $V=c(1+\varepsilon)$  and the delay is small, one has $r \simeq 2 + {\varepsilon ^2}$  and  $s \ll 1$; but we shall also consider extreme cases of large superluminality ($r \gg 1$) and large delay or very small flight time ($s \gg 1$).

Eq.\ (\ref{eqR4}) is rewritten as
\begin{equation}
2 - \bar \beta r + s\sqrt {1 - {{\bar \beta }^2}}  = 0
\label{eqR6} 
\end{equation}
First we show in an elementary way through series expansions that with a signal which is only slightly superluminal ($V=c(1+\varepsilon)$) we can compensate for any delay and we always obtain an active causality violation for some $\bar \beta$. In fact, substituting into (\ref{eqR6}) the Ansatz
\begin{equation}
\bar \beta  = 1 - \frac{1}{2}\eta {\varepsilon ^2},{\rm{  \ \     with    \ \  }}0 < \eta  < 1
\label{eqR7} 
\end{equation}
we obtain (for $\eta  \to 0$  and $\varepsilon$ small)
\begin{equation}
s=\frac{{{\varepsilon ^2}\left( {1 - \eta } \right) + g\left( {{\varepsilon ^2}} \right)}}{{\varepsilon \sqrt \eta  \left( {1 - \frac{1}{8}\eta {\varepsilon ^2} + h\left( {\eta {\varepsilon ^2}} \right)} \right)}} \approx \frac{{\varepsilon \left( {1 - \eta } \right)}}{{\sqrt \eta  }}
\label{eqR8} 
\end{equation}
(The functions $g$ and $h$ are at least quadratic.)

We see that by choosing $\eta$ close to zero, we can obtain a value of $s$ as large as we want. This means that, in reverse, it is always possible to find a solution for  $\bar \beta$, no matter how large the delay $s=\Delta t/t_B$.

In order to find the general solution of (\ref{eqR6}) and examine it in the strongly superluminal case $V \gg c$, $r \gg 2$, we transform (\ref{eqR6}) into a quadratic equation. By taking the square on both sides, we find that the solution of this equation is also solution of the following equation
\begin{equation}
{\bar \beta ^2}\left( {{r^2} + {s^2}} \right) - 4\bar \beta r + 4 - {s^2} = 0
\label{eqR9} 
\end{equation}
(There is also one spurious solution, see below.)

The discriminant $\Delta  = 4{s^2}\left( {{s^2} + {r^2} - 4} \right)$  is always positive. The solutions of (\ref{eqR9}) are
\begin{equation}
{\bar \beta _{1/2}} = \frac{{2r \pm s\sqrt {{s^2} + {r^2} - 4} }}{{{r^2} + {s^2}}}
\label{eqR10} 
\end{equation}
but it is straightforward to check that only the solution with the plus sign is acceptable as a solution of (\ref{eqR6}). The check is most readily done in the limit $s \ll r$. In this same limit, we can also rewrite (\ref{eqR10}) as
\begin{equation}
\bar \beta  \approx \frac{2}{r} + \frac{s}{r}{\rm{ \ \     (for \   }}s \ll r)
\label{eqR11} 
\end{equation}
and we see directly that $0 < \bar \beta  < 1$, as it should be. From eq.\ (\ref{eqR11}) we also see that when $r$ is large and $s$ is approximately 1 (delay approximately equal to propagation time), one has $\bar \beta  \simeq 3/r$.

\begin{figure}
  \includegraphics[width=9cm,height=6cm]{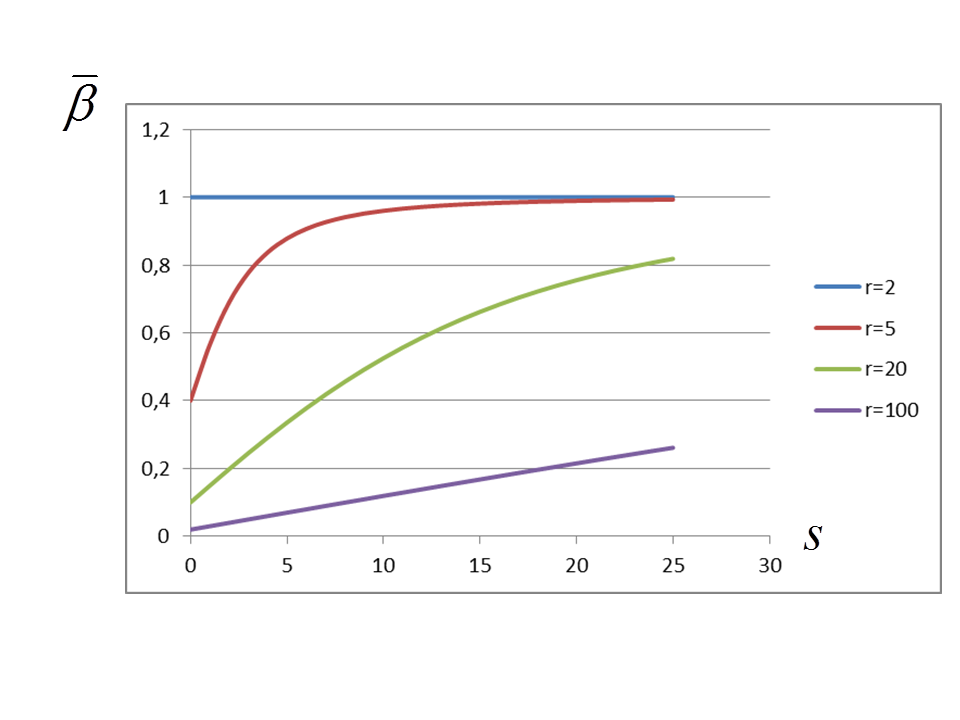}
\caption{Velocity requirement for active violation in the fairly superluminal case with large delay ($r>2$, $s$ up to 25).}
\label{Fig5}       
\end{figure}

The fairly superluminal case with large delay ($r>2$, $s \gg 1$) or strongly superluminal with large delay ($r\gg 2$, $s \gg 1$) are more difficult to study in an analytical way. By plotting  $\bar \beta $ as a function of $r$ and $s$ (Fig.\ \ref{Fig5}) one can check that the solution of (\ref{eqR10}) with the plus sign always gives a value of $\bar \beta $  in the range $0<\bar \beta <1$, and that $\bar \beta  \to {1^ - }$  when $s \to  + \infty $ . This means that it is always possible to obtain an active causality violation, compensating any delay in the secondary emission, provided the moving emitter is travelling fast enough. (Having already shown this in eq.\ (\ref{eqR8}) in the case of slightly superluminal velocity, we should not be surprised to find a general confirmation for strongly superluminal signals.)

\section{Discussion. Delay due to precursor propagation}

The conclusion that active violation is possible also in the presence of an arbitrary delay is mathematically clear and is a consequence of the singularity of the Lorentz transformation for  $ \beta  \to {1^ - }$. (Such a singularity could be eliminated through a physical cut-off, based for instance on the existence of a maximum acceleration for reference frames \cite{Toller,Lambiase}). Physically, the situation looks puzzling. Take, for instance, the following extreme case: we have a primary superluminal signal that travels for a small fraction of a second and hits a receiver, which triggers a moving secondary emitter; a secondary tachyon is emitted with delay of, say, {\it one year}, and travels back to the primary emitter. Note that the delay is measured in the moving system and will appear even longer in the lab system. And yet there exist moving systems such that the secondary tachyon still can reach the primary emitter before the primary emission! The intuitive reason is that in the moving systems, when $ \beta  \to {1^ - }$, the propagation time $t_B'$ appears to be very long, and eventually much longer than any fixed delay $\Delta t$.

\begin{figure}
  \includegraphics[width=9cm,height=6cm]{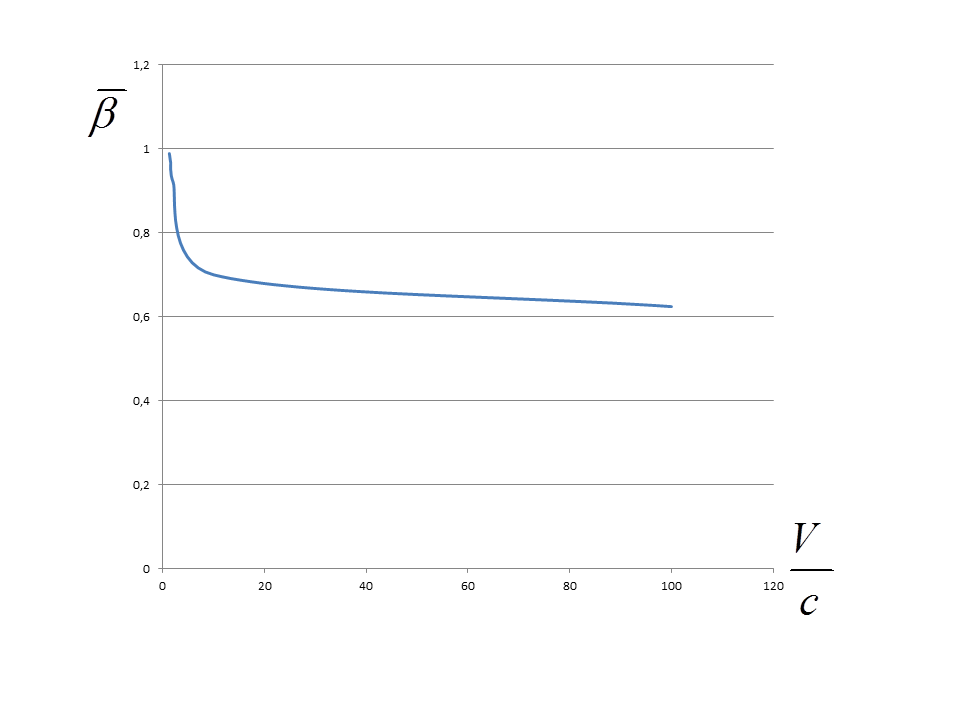}
\caption{Velocity requirement for active causality violation in the presence of a delay due to precursor propagation (numerical solution of eq.\ (\ref{numerical})). Compare with Fig.\ \ref{beta}.}
\label{beta-new}       
\end{figure}

Let us now evaluate the requirement for active violation supposing that the delay in the secondary emission is due to the propagation of a luminal precursor before the superluminal signal. As mentioned in the Introduction, it has been suggested that such precursors must be present in superluminal signals and prevent them from really transferring information at superluminal speed. In the moving frame, it takes a time $\Delta t = x_B/(\gamma c) = V t_B/(\gamma c)$ for the precursor to reach the primary emitter. Denoting $\xi=V/c$, this means that the parameter $s$ introduced in eq.\ (\ref{eqR5}) becomes $s=\xi/\bar \gamma$, where $\bar \gamma$ is the critical value $\bar \gamma=(1-\bar \beta)^{-1/2}$. By inserting this into (\ref{eqR10}) we find the following equation in $\xi$, $\bar \gamma$:
\begin{eqnarray}
 & &\sqrt{1-\frac{1}{\bar \gamma}} \left[ \left( \frac{\xi}{\bar \gamma} \right)^2 + \left( \xi + \frac{1}{\xi} \right)^2 \right] = \label{numerical}\\
& &= 2 \left( \xi + \frac{1}{\xi} \right) +  \frac{\xi}{\bar \gamma} \sqrt{\left( \frac{\xi}{\bar \gamma} \right)^2 + \left( \xi + \frac{1}{\xi} \right)^2 -4 }
\nonumber
\end{eqnarray}
This can be solved numerically in $\bar \gamma$, giving the graph of Fig.\ \ref{beta-new}. The result is an enhancement of the ``unrealistic'' requirements for the small values of $V/c$ reported for X-shaped waves: for instance, for $\xi=V/c=1.25$ we have $\bar \beta = 0.989$. Moreover, eq.\ (\ref{numerical}) sets a lower limit on the requirements for extreme superluminal signals: when $\xi \to \infty$, $\beta \geq 0.62$ (to be compared with Fig.\ \ref{beta}, where $\bar \beta \to 0$ when $\xi \to \infty$).

\section{Conceptual problems in interactions via tachyon exchange}
\label{conceptual}

\begin{figure}
  \includegraphics[width=9cm,height=6cm]{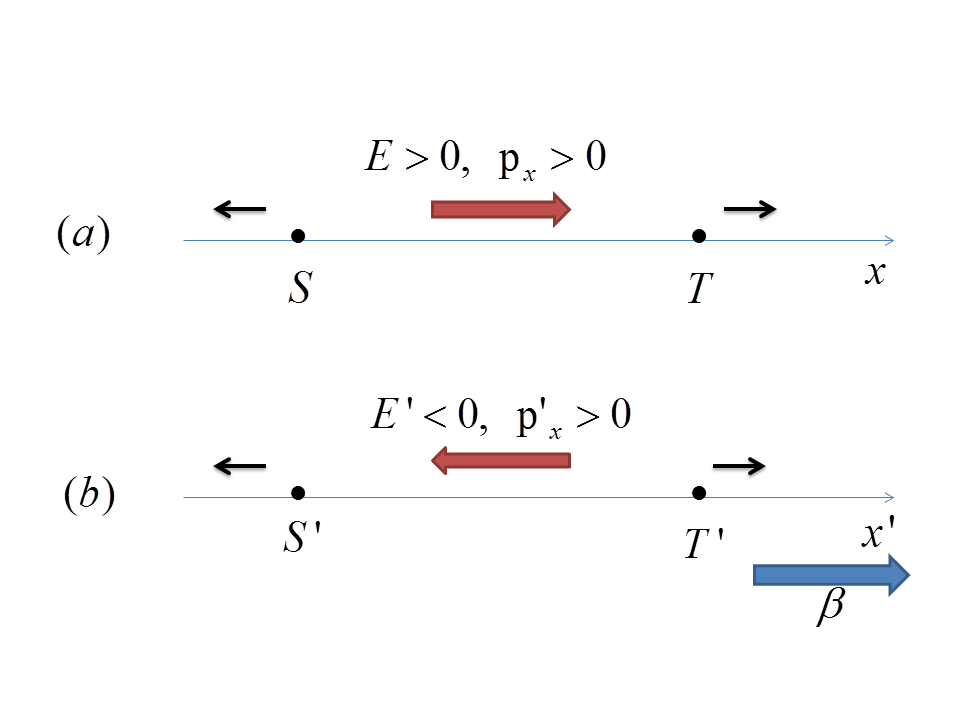}
\caption{Trying to distinguish source (S) from target (T) in a tachyon exchange leads to inconsistencies. In (a), S is seen to emit a tachyon, while T absorbs it. The tachyon carries a positive energy and a positive momentum from S to T (big red arrow). S recoils to the left, T recoils to the right (black small arrows). (b) The same process as observed in a moving frame with passive causality violation. The tachyon appears to be emitted from T' and absorbed by S' at a later time (big red arrow). According to the Lorentz transformation (\ref{eq5-7}) the tachyon carries a negative energy, thus the total energetic balance of the process is the same as observed in the rest system. The recoil momenta have the same sign as before, but the transferred momentum is still positive. Therefore the ``direction'' of the momentum exchange appears to be incompatible with the recoil, in the sense that after T' has emitted some positive momentum, it should not recoil to the right, and similarly for S'. }
\label{recoil}       
\end{figure}

\begin{figure}
  \includegraphics[width=9cm,height=6cm]{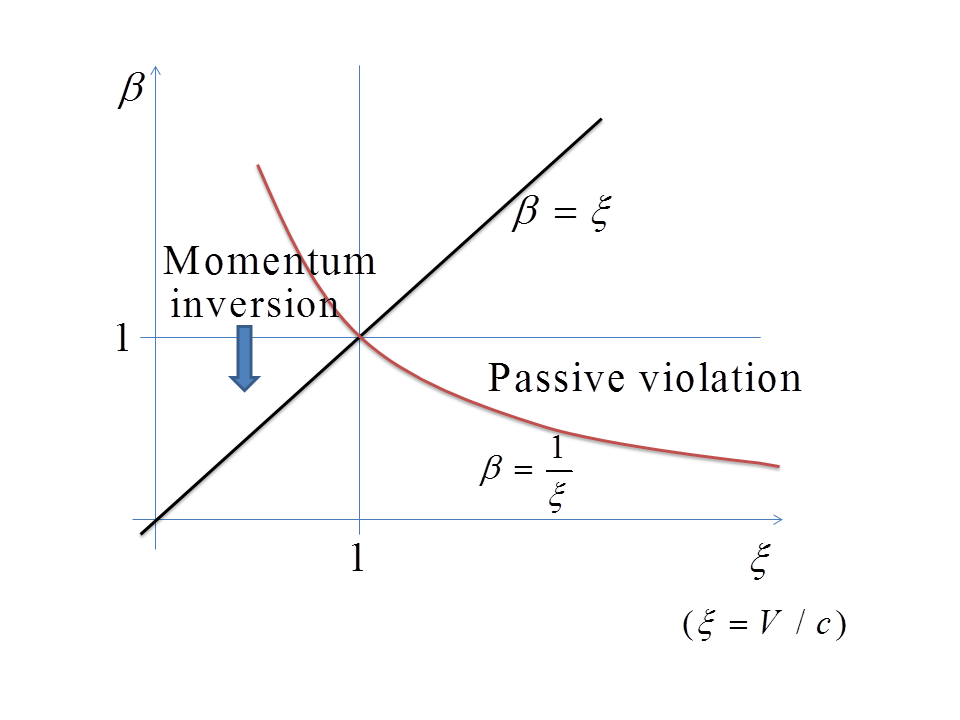}
\caption{Proof that conditions for passive violation and momentum inversion are incompatible. On the horizontal axis, $\xi$ represents the ratio between tachyon velocity and light velocity. The range of interest for $\xi$ is $\xi>1$. On the vertical axis, $\beta$ is the parameter of a Lorentz boost ($0<\beta<1$). Passive causality violation of the tachyon is observed for  $\beta>c/V$, therefore for $\beta>1/\xi$, which is the region of the graph between the lines $\beta=1/\xi$ and $\beta=1$. From eq.\ (\ref{eq5-7}) we see that the momentum of the tachyon, as observed from the moving system, changes sign if $1-\beta c/V<0$, i.e.\ if $\beta > \xi$. The latter is the region between the lines $\beta=\xi$ and $\beta=1$, which does not have any point in common with the region of passive causality violation.  }
\label{incompatible}       
\end{figure}

In this section we consider a process in which two bodies interact through the exchange of short-lived superluminal particles. For convenience, we shall refer again to these particles as tachyons, with the understanding that they cannot exist as real stable particles. According to the discussion of the preceding sections, we disregard the problems related to the possibility of active causal violation. Yet some further conceptual problems are encountered when one tries to describe the exchange interaction as a sequence of tachyon emission-propagation-absorption. If we assume that one of the bodies emits the tachyons and the other is the target, their roles are clearly exchanged in a system in relative motion where time reversal is observed, and the energy of the tachyons change sign accordingly; but their momentum does not change sign, preventing a consistent re-interpretation of the process. This is a straightforward though subtle consequence of the Lorentz transformations, summarized in Fig.s \ref{recoil}, \ref{incompatible} and their captions. 

The logical consequences appear to us to be the following: it is inconsistent to say that in any circumstance two bodies interact through the exchange of tachyons ``emitted'' from one of the two bodies and ``absorbed'' by the other. Actually, if two bodies interact through the exchange of tachyons, it is impossible to say which body emits them and which one absorbs them, and to say that at some time the tachyon is carrying energy $E$ and momentum $p$ from one body to the other. We can only consider the exchange process as a whole. (This should not be surprising if one recalls the failure of the local realism principle in typical quantum processes like the EPR phenomenon.)

All this implies that if conservation laws impose any constraints on the exchange either in the ``source'' or in the ``target'' in anyone of the reference frames,
this constraint will apply even if (apparently) backwards in time. For instance, if a source emits a tachyon which is elastically absorbed by a non-relativistic target particle T with energy-momentum ratio 
\begin{equation}
\frac{E_T}{p_T}=\frac{\frac{1}{2}mv_T^2}{m_Tv_T}=\frac{1}{2}v_T 
\end{equation}
then the $E/p$ ratio of the tachyon must be equal to $\frac{1}{2}v_T$ and therefore ``the target determines the propagation velocity of the tachyon and the recoil of the source''. (Again, this looks reasonable if one regards the interaction process as a whole quantum process.)

For the proofs we only need a few basic relations concerning kinematics and Lorentz transformations. Extensions of Special Relativity exist, compatible with the relativity principle, which provide a complete framework for the kinematics of superluminal particles (\cite{break} and ref.s). The usual definitions of energy and momentum are extended by introducing an imaginary mass $m=iM$:
\begin{equation}
E = \frac{{m{c^2}}}{{\sqrt {1 - {V^2}/{c^2}} }}=
\frac{{M{c^2}}}{{\sqrt {{V^2}/{c^2}-1} }}
\label{eq5-3} 
\end{equation}
\begin{equation}
{\bf p} = \frac{{m{\bf V}}}{{\sqrt {1 - {V^2}/{c^2}} }}=
\frac{{M{\bf V}}}{{\sqrt {{V^2}/{c^2}-1} }}
\label{eq5-4} 
\end{equation}
Combining eq.s (\ref{eq5-3}) and (\ref{eq5-4}) one obtains
\begin{equation}
{E^2} - {p^2}{c^2}={m^2}{c^4}  < 0
\label{eq5-1} 
\end{equation}
\begin{equation}
\frac{{\bf p}}{E} = \frac{{\bf V}}{{{c^2}}}
\label{eq5-5} 
\end{equation}
Note that if $E<0$ (tachyon observed from a system with passive violation, see below), then $M<0$, and ${\bf V}$ and ${\bf p}$ are opposite.

The Lorentz transformations of the energy and momentum of the exchanged tachyon for a boost in the $x$ direction are
\begin{equation}
\begin{array}{l}
E' = \gamma \left( {E - \beta cp_x} \right)\\
p'_x = \gamma \left( {p_x - \beta \frac{E}{c}} \right)
\end{array}
\label{eq5-6} 
\end{equation}
Suppose that the momentum is directed along $x$.
Remembering that $p_x/E = V_x/{c^2}$ and denoting the positive value of $V_x$ simply as $V$, one has
\begin{equation}
\begin{array}{l}
E' = \gamma E\left( {1 - \beta \frac{V}{c}} \right)\\
p'_x = \gamma p_x\left( {1 - \beta \frac{c}{V}} \right)
\end{array}
\label{eq5-7} 
\end{equation}

Suppose to be in the case of passive violation, i.e.\ to observe the tachyon from a moving reference frame with velocity $v/c>c/V$ (Fig.\ \ref{recoil}). We see from (\ref{eq5-7}) that $E'<0$. This is consistent with the inversion of time in the tachyon propagation, in the following sense: in the lab frame, the tachyon is seen to transfer a positive energy from the source to the target; in the moving frame we see the tachyon leave the target and carry away a negative energy which is delivered to the source at a later time; thus the final energetic balance is the same, in the sense that some energy has passed from the source to the target. Unfortunately, however, a similar reasoning does not hold for the momentum transfer, because from eq.\ (\ref{eq5-7}) we see that in the moving frame $p'>0$, like in the laboratory. (See Fig.\ \ref{incompatible}.) Therefore it is inconsistent to assume that there is a source and a target in the process, even admitting that they are exchanged in certain Lorentz trasformations.


\begin{thebibliography}{00}



\bibitem{eva1}
{Steinberg A.M., Kwiat P.G. and Chiao R.Y.},
   {Phys. Rev. Lett.}{\ 71}{\ (1993)\ }{708}.

\bibitem{eva2}
   {Enders A.  and Nimtz G.},
   {J. Phys. I France}{\ 2}{\ (1992)\ }{1693}.

\bibitem{eva3}
   {Mugnai D., Ranfagni A.  and Ronchi L.},
   {Phys. Lett. A}{\ 247}{\ (1998)\ }{281}.

\bibitem{eva4}
   {Balcou P.  and Dutriaux L.},
   {Phys. Rev. Lett.}{\ 78}{\ (1997)\ }{851}.

\bibitem{Mugnai}
   {Mugnai D., Ranfagni A. and Ruggeri R.},
   {Phys. Rev. Lett.}{\ 84}{\ (2000)\ }{4830}.

\bibitem{Opt1}
   {Saari P. and Reivelt K.},
   {Phys. Rev. Lett.}{\ 79}{\ (1997)\ }{4135}.

\bibitem{Opt2}
   {Durnin J., Miceli J.J. and Eberly J.H.},
   {Phys. Rev. Lett.}{\ 58}{\ (1987)\ }{1499}.

\bibitem{miss}
   {Missevitch O.V., Kholmetskii A.L. and  Smirnov-Rueda R.}, 
   {Europhys. Lett.}{\ 93}{\ (2011)\ }{64004}.

\bibitem{break}
   {Chashchina O.I. and  Silagadze Z.K.}, 
   {Acta Phys. Pol. B}{\ 43}{\ (2012)\ }{1917}.

\bibitem{Benford}
   {Benford G., Book D.  and  Newcomb W.}, 
   {Phys. Rev. D}{\ 2}{\ (1970)\ }{263}.

\bibitem{Toller}
   {Toller M.}, 
   {Int. J. Theor. Phys.}{\ 29}{\ (1990)\ }{963}.

\bibitem{Lambiase}
   {Nesterenko V.V., Feoli A., Lambiase G. and Scarpetta G.}, 
   {Phys. Rev. D}{\ 60}{\ (1999)\ }{065001}.




\end{thebibliography}
\end{document}